\documentclass[prb,twocolumn,showpacs,superscriptaddress,amsmath,amssymb]{revtex4}
\usepackage{mathrsfs}

\usepackage{graphicx}
\usepackage{amssymb}
\usepackage{dcolumn}
\usepackage{bm}
\usepackage{color}

\begin{document}


\title{Manipulating the fully spin-polarized edge currents in graphene ribbon}

\author{Lei Xu}
\affiliation{National Laboratory of Solid State Microstructures and
Department of Physics, Nanjing University, Nanjing 210093, China}

\author{Jin An}
 \email{anjin@nju.edu.cn}
\affiliation{National Laboratory of Solid State Microstructures and
Department of Physics, Nanjing University, Nanjing 210093, China}

\author{Chang-De Gong}
\email{cdgongsc@nju.edu.cn}
\affiliation{Center for Statistical and Theoretical Condensed Matter
Physics, and Department of Physics, Zhejiang Normal University,
Jinhua 321004, China}

\affiliation{National Laboratory of Solid State Microstructures and
Department of Physics, Nanjing University, Nanjing 210093, China}

\date{\today}

\begin{abstract}
Electron fully spin-polarized edge states in graphene emerged at the interfaces of a nonuniform magnetic field are studied numerically in a
tight-binding model, with both the orbital and Zeeman-splitting effects of magnetic field considered. We show that the fully spin-polarized
currents can be manipulated by a gate voltage. In order to make use of the fully spin-polarized currents in the spin related transport, a
three-terminal experiment is designed and expected to export the fully spin-polarized currents. This may have important applications in spin
based nanodevices.
\end{abstract}

\pacs{72.25.-b, 73.20.-r, 71.70.Di}

\maketitle

\section{introduction}
Recently, spin-dependent transport phenomena have received intensive studies because of their potential applications in spintronics devices.
Understanding how to generate and manipulate the spin-polarized currents is one of the key points for the development of the spin-based devices.
Rashba spin-orbit interaction, which can be controlled by an (or effective) electric field, provides an efficient way to achieve a
spin-polarized current, for example in graphene ribbon,\cite{Zarea2009} doped semiconductors\cite{Kato2004} and 2D electron
system.\cite{Sinova2004} In most of these applications, the induced charge currents are only partially polarized,\cite{Kane2005} but a fully
spin-polarized (FSP) current is much more difficult to achieve experimentally. Up to now, FSP current is found to be realized only in very
limited materials, for example, in the half-metallic states of the ferromagnetic metals\cite{FerroMetA,FerroMetB} and ferromagnetic
semiconductors,\cite{FerroSemA} where the electronic valance band is partially filled for one spin component, whereas completely empty for the
other. From material viewpoint, two-dimensional (2D) graphene, as a candidate material of new generation electronics\cite{Castro
Neto2009,Geim2007,Meyer2007} for its high mobility and low carrier density, is expected to have great application in spintronics
devices.\cite{Cho2007,Tombros2007,Hill2007,Lundeberg2009} How to realize a FSP current in graphene is then quite meaningful for future
graphene-based electronic devices. In graphene nanoribbon with zigzag edges, it has been predicted from a fist-principle
calculation\cite{Louie2007} and numerical simulations\cite{Mac2009a,Mac2009b} that with the application of a transverse electric field, a
half-metallic state can be achieved, which leads naturally to FSP edge currents propagating along sample boundaries.

On the other hand, for a 2D electron system, a sufficient strong
magnetic field will polarize the spins of all electrons in the
same direction as the field, but only a much smaller magnetic
field is needed to achieve a FSP current.\cite{Xu2010} The cost is
that a nonuniform magnetic field (NMF) must be applied to the
system. Actually, it has been predicted that a partially
spin-polarized current will appear at the interfaces of
inhomogeneous magnetic fields,\cite{RamezaniMasir2008,Park2008}
but much less attention are taken on the FSP states at the
interfaces, especially, in graphene.

\begin{figure}
\scalebox{0.7}[0.7]{\includegraphics[150,512][442,701]{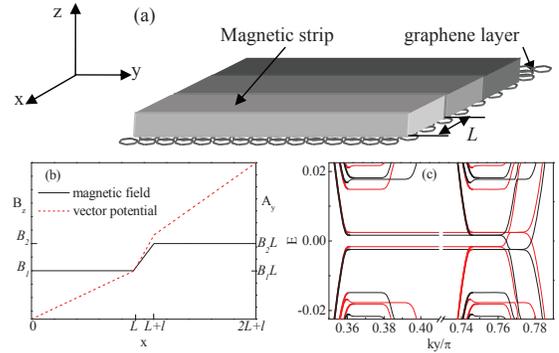}}
\caption{\label{fig1}(Color online) (a) Schematic illustration of
the system: an array of ferromagnetic strips with different
magnetizations is placed on the top of a rectangular graphene
layer. In the two-strip case, (b) the magnetic field (solid line)
and its gauge potential (dotted line) as functions of $x$, and (c)
the corresponding energy spectrum for a zigzag-edge graphene
ribbon with $L=213$~nm, $l=2.13$~nm, $B_1=2$~T and $B_2=-3$~T.}
\end{figure}

In this paper, we study monolayer graphene in the presence of a
stepwise NMF, which can be realized experimentally in several
ways,\cite{Cerchez2007,Leadbeater1995,Bending1990} to exhibit
systematically how to manipulate the FSP currents generated at the
interfaces of the NMF. The situation is schematically shown in
Fig.~\ref{fig1}(a). For the uniform magnetic field case, the FSP
state is found to be absent, and the transport in graphene can be
described by the so-called counter-propagating chiral edge
states,\cite{Abanin2006} when the chemical potential $\mu$ is
between the spin gap due to Zeeman splitting. However, for the
stepwise NMF case, several spin gaps are opened and the FSP edge
states may emerge at the magnetic interfaces when $\mu$ is between
the gaps. In the following we will address the approach to
manipulate this novel FSP edge currents. Furthermore, an
experimental setup is designed to export the edge currents.

\section{model}
We begin our discussion by the tight-binding model on a honeycomb
lattice in the presence of a perpendicular magnetic field
$\mathbf{B}=(0,0,B_z(x))$, which is given by,
\begin{equation}
H_0=-t\sum_{<ij>}[e^{i\int_i^j\mathbf{A}\cdot d\mathbf{l}}\hat{c}_{i}^\dag \hat{c}_{j}+ H.c.]-\sum_i h_Z(x)(n_{i\uparrow}-n_{i\downarrow}),
\label{eq:one}
\end{equation}
where the first term describes electrons hopping between the nearest neighbors, suffering an additional phase caused by the orbital effect of
magnetic field, with $\hat{c}_i^\dag=(\hat{c}_{i\uparrow}^\dag,\hat{c}_{i\downarrow}^\dag)$ electron creation operators, whereas the second one
the Zeeman splitting term. Here $h_Z(x)=\frac{1}{2}g\mu_BB_z(x)$ with $g$ as the Land\'{e} factor. The magnetic field $B_z(x)$ is assumed to
change only in $x$ direction, and the stepwise one which is focused on in this paper, has the following form,
\begin{equation}
B_z(x)=
\begin{cases}
B_1& 0<x<L,\\
B_1+\frac{B_2-B_1}{l}(x-L)& L<x<L+l,\\
B_2& L+l<x<2L+l.
\end{cases}\label{eq:B}
\end{equation}
For the gauge potential $\mathbf{A}$, the Landau gauge $\mathbf{A}=(0,A_y(x),0)$ is adopted, with $A_y(x)=\int_0^xB_z(x')dx'$ given as:
\begin{equation}
A_y(x)=
\begin{cases}
B_1x& 0<x<L,\\
B_1x+\frac{B_2-B_1}{2l}(x^2-L^2)& L<x<L+l,\\
B_2x-\frac{B_2-B_1}{2}l& L+l<x<2L+l.
\end{cases}\label{eq:A}
\end{equation}
For simplicity, in the following discussion, energy is measured in unit of the hopping integral $t$.

\section{results and discussions}

To analyze all possible edge states, we choose open (periodic)
boundary condition in the $x$ ($y$) direction, and numerically
diagonalize the Hamiltonian $H_0$ on a rectangular sample under a
NMF to obtain the electron energy spectrum. In one interface case,
the spectrum is shown in Fig.~\ref{fig1}(c), which is symmetric
with respect to zero energy due to the particle-hole symmetry (
which, strictly speaking, should be combined with the
time-reversal symmetry) preserved even in a NMF. The spectrum is
composed of two series of Landau levels (LLs), including all the
bulk states of the system, and branches of edge states, which play
a dominant role in quantum transport in the presence of magnetic
field. The edge states can be classified into the normal ones
located at the sample boundaries, and the interface ones located
at the interface where the magnetic field changes abruptly.

\subsection{Generation and manipulation of the FSP edge currents}

\begin{figure}
\scalebox{0.65}[0.65]{\includegraphics[116,328][459,710]{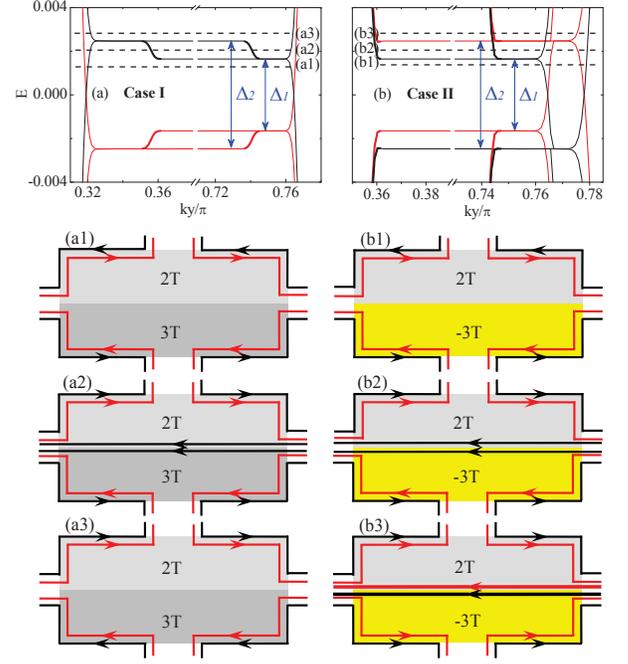}}
\caption{\label{fig2}(Color online) Top panels: Electron energy
spectrum of graphene ribbon under a stepwise magnetic field with
one interface, with $L=213$~nm (1000 zigzag chains) for (a) Case
\textbf{I} with $B_1=2$~T and $B_2=3$~T, (b) Case \textbf{II} with
$B_1=2$~T and $B_2=-3$~T. Here the transition length for the
variation of magnetic field long $x$ direction is $l=2.13$~nm (10
zigzag chains). The red and black lines denote bands of spin-up
and spin-down states, respectively. The interface edge states are
labelled with thick lines. The Zeeman energy is chosen to be one
tenth of the largest LL spacing of the region with $B=2$~T. The
black dashed lines give the representative positions of the
chemical potentials $\mu$. Other panels: The corresponding
four-terminal configurations in Hall measurement. The spatial
distribution of the edge currents is shown, where the red and
black lines with arrows represent the spin-up and spin-down
electron edge currents, respectively. The thick lines with arrow
at the interfaces in (b3) represent that each current is carried
by two edge states.}
\end{figure}

Let's focus our discussion on the regime where the chemical
potential $\mu$ is between the zeroth LLs of the two spin species.
To make things more explicit, we show in Figs.~\ref{fig2}(a) and
\ref{fig2}(b) the energy spectrum around zero energy. The
splitting of the zeroth LL and formation of two spin gaps
$\Delta_1$ and $\Delta_2$ can be clearly seen. When $\mu$ lies
within the interval $|\mu|<\Delta_1/2$ or $|\mu|>\Delta_2/2$,
there exist only sample boundary edge states, and no interface
edge states at all. However, when $\mu$ lies between the two spin
gaps $\Delta_1/2<|\mu|<\Delta_2/2$, interface edge states are
generated.

\begin{figure*}
\scalebox{0.8}[0.8]{\includegraphics[3,303][588,741]{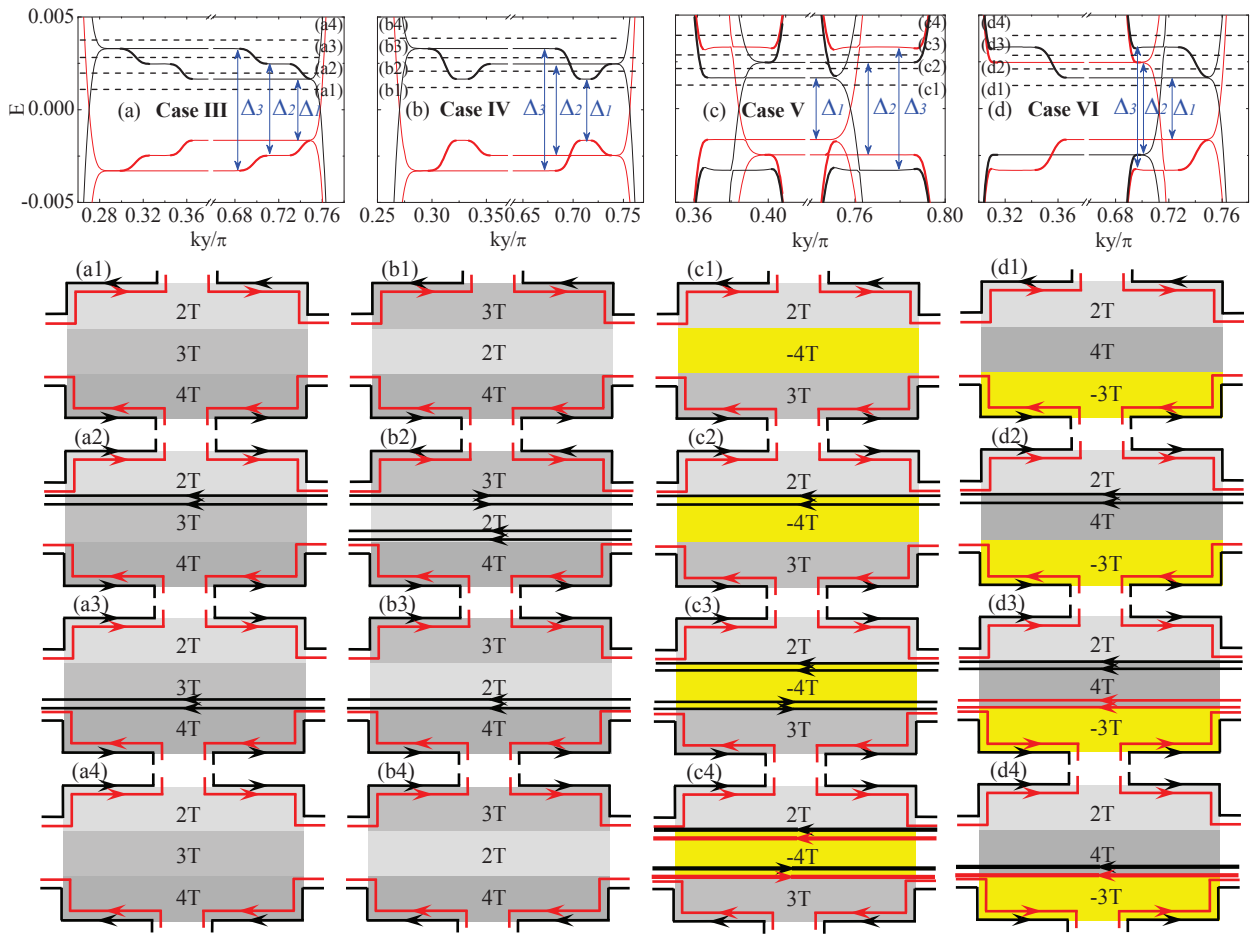}} \caption{\label{fig3}(Color online) Similar to that of Figs.~\ref{fig2}, but for
four representative configurations of two interfaces (a) Case \textbf{III} with $B_1=2$~T, $B_2=3$~T and $B_3=4$~T, (b) Case \textbf{IV} with
$B_1=3$~T, $B_2=2$~T and $B_3=4$~T, (c) Case \textbf{V} with $B_1=2$~T, $B_2=-4$~T and $B_3=3$~T, and (d) Case \textbf{VI} with $B_1=2$~T,
$B_2=4$~T and $B_3=-3$~T.}
\end{figure*}

For $0<\mu<\Delta_1/2$, edge currents are propagating in opposite
directions for opposite spin polarizations at both sample
boundaries [see Fig.~\ref{fig2}(a1)], which is similar to the
quantum spin Hall state in graphene by considering intrinsic SO
interaction,\cite{Kane2005} and also similar to the state
predicted in Ref.\onlinecite{Abanin2006} in a uniform magnetic
field. For $\mu>\Delta_2/2$, however, the edge currents at sample
boundaries become normal because they propagate in the same
direction for both spin polarizations [see Fig.~\ref{fig2}(a3)].
Interestingly, when the chemical potential $\mu$ lies within the
gap $\Delta_1/2<\mu<\Delta_2/2$, the interface channels are opened
and two spin-down edge states with the same flowing direction
emerge at the interface, and certainly, they are FSP
[Fig.~\ref{fig2}(a2)]. Meanwhile, the edge current at one sample
boundary becomes normal whereas that at the other boundary is
still left novel. For case where the magnetic fields in the two
sides of the interface are in opposite directions, similar
conclusion can be made except that when $\mu>\Delta_2/2$, no FSP
state is present but the interface current channel is not closed
and there exist a normal edge current at the interface instead
[see Figs.~\ref{fig2}(b1)-\ref{fig2}(b3)].

From the viewpoint of application, one can use the properties
mentioned above to manipulate the FSP edge currents. Imaging
placing a voltage gate on the top of graphene sample, and by
changing gradually the voltage, one can vary the carrier density
in the sample, so the chemical potential $\mu$ gradually. We take
Case \textbf{I} as an example. When $\mu$ lies within the minimum
spin gap $|\mu|<\Delta_1/2$, the interface edge current channel is
closed. As $|\mu|$ increases, the interface channel with a FSP
edge current is opened if $\mu$ lies within the two gaps
$\Delta_1/2<|\mu|<\Delta_2/2$. Finally the channel is closed again
in the interval $|\mu|>\Delta_2/2$. This implies that one can
control the FSP edge currents electrically.

The spin gap due to Zeeman splitting is estimated as $\sim 3$~K at $B=2$~T, whereas the corresponding first LL $E_1$ is estimated to be about
500K. Furthermore, it was argued that the spin gap can be enhanced to a few hundred kelvin for a realistic magnetic field due to electron
exchange interaction.\cite{Abanin2006} So the FSP current is expected to be observed experimentally in relatively large range of temperatures.
To visualize the effect of the Zeeman splitting clearly and not changing the physics, relatively large spin gaps have been chosen for
calculating the energy spectrum in Fig.~\ref{fig2}. We also note that the localization character along $x$ direction for the edge states is well
defined because the magnetic length here is $l_B=\sqrt{\hbar c/eB}=18$~nm, much less than the width of the sample $L$ (=213~nm).

We next consider the case of NMF with two interfaces. There are
twelve in total different configurations for pattern of magnetic
field, but among them, only four( Case \textbf{III}- Case
\textbf{IV}) are representative and of great interest, which is
shown in Fig.~\ref{fig3}. Three spin gaps are formed in each case,
so similar to Fig.~\ref{fig2}, the states between the zero LLs can
be generally classified into four regimes, according to the
position of the chemical potential $\mu$.

Case \textbf{III} is very special. For this interesting case, if
generally, there are more interfaces with the magnetic field in
each region in the same direction and increasing monotonically,
the FSP interface edge channels will be opened one by one from the
low-field interface to high-field one, where the old one closing
and new one opening happen simultaneously. For Case \textbf{IV},
when $\mu$ increases gradually from $0<\mu<\Delta_1/2$ to
$\mu>\Delta_3/2$, the two interface channels are opened
simultaneously, after that one channel is closed and finally the
other one is closed. For Case \textbf{V} and Case \textbf{VI},
although the FSP interface channels are opened and closed in
different orders, these channels are not closed in the end but
become normal ones instead [Figs.~\ref{fig3}(c4) and
\ref{fig3}(d4)].

Based on the above discussion, we come to a general conclusion
that for a NMF with many interfaces, so long as the magnetic
length is much less than the width of each uniform region, there
always exist FSP edge states either at all the interfaces or at
several of them, when the chemical potential $\mu$ lies within the
gap $\Delta_{min}/2<|\mu|<\Delta_{max}/2$. All the calculations
above are performed for the zigzag graphene ribbon. For the
armchair graphene ribbon, similar edge states and so similar
conclusion can be obtained.

\subsection{Rashba spin-orbit coupling effect}

\begin{figure}
\scalebox{0.7}[0.7]{\includegraphics[101,628][435,754]{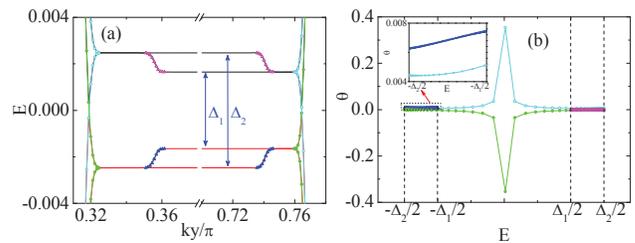}}
\caption{\label{fig4}(Color online) (a) Electron energy spectrum
of graphene in the presence of Rashba SO interaction for Case
\textbf{I}. Parameters are the same as that in Fig.~\ref{fig2} (a)
and $\lambda=\Delta_1/50$. (b) The spin tilt angle $\theta$ as a
function of $E$ for different edge states. The inset in (b) is an
enlarged view of the tilt angle within the spin gap
$-\Delta_2/2<\mu<-\Delta_1/2$. Open and filled triangles denote
the interface edge states, while open and filled circles denote
the sample boundary edge states.}
\end{figure}

We now discuss the effect of the spin-flip processes caused by
spin-orbit(SO) interaction. To estimate the influence of the SO
interaction on the polarized edge current, we only consider here
the Rashba SO interaction as an example, whose Hamiltonian can be
written as
\begin{equation}
H_R=\lambda\sum_{<ij>}[ie^{i\int_i^j\mathbf{A}\cdot d\mathbf{l}}\hat{c}_{i}^\dag (\mathbf{\sigma}\times \mathbf{d}_{ij})\cdot \mathbf{\hat{z}}
\hat{c}_{j}+H.c.], \label{eq:R}
\end{equation}
where $\lambda$ is the Rashba SO coupling strength and
$\mathbf{\sigma}$ is the Pauli matrix. $\mathbf{d}_{ij}$ is a
vector pointing from $j$ to its nearest neighbor $i$, and
$\mathbf{\hat{z}}$ is a unit vector in $z$ direction. Figure
\ref{fig4} (a) shows the spectrum at low energies with Rashba SO
interaction, which is found to be almost unchanged compared with
the case without SO coupling [Fig.~\ref{fig2} (a)]. However, in
contrast to the case without SO interaction where electrons are
polarized along $z$ axis, the spin polarization of the edge states
in the presence of the SO interaction can be described by
$\mathbf{S}=(\sin\theta,0,\cos\theta)$ with $\theta$ the tilting
angle deviated from the z axis, changing continuously with $k_y$.
Therefore, theoretically speaking, the FSP interface current is
still preserved at $T=0$~K [Fig.~\ref{fig4} (a)], except that the
edge current is now polarized along other direction instead which
slight deviates from $z$ axis by an angle $\theta(k_F)$. In the
spin gap region $\Delta_1/2<|\mu|<\Delta_2/2$, for the parameters
shown in Fig.4, the tilting angle for the interface edge states is
estimated to be $|\theta|<0.0074$. It is very small and changes
nearly linearly with energy with the slope $\Delta\theta/\Delta
E\sim1.3$ in this gap [see the inset in Fig.~\ref{fig4} (b)].
Physically, the slope $\Delta\theta/\Delta E$ is believed to be
nearly proportional to $\lambda$, which is rather small in real
situations (For a perpendicular electric field $E_z\sim
50$V$/300$~nm,\cite{Kane2005,Novoselov2004} $\lambda\sim
0.07$~K.\cite{Huertas2006,Min2006}). At finite temperatures, in
order to obtain a almost FSP interface current, the
thermodynamical energy $k_BT$ should be much less than the spin
gap $(\Delta_2-\Delta_1)/2$, which is estimated to be about $30$~K
for $B=2$~T.\cite{Abanin2006} Therefore, in many situations
achievable in experiments, the spin-flip processes induced by the
Rashba SO interaction can be negligible, and an almost FSP current
can be realized.

\subsection{Exporting the FSP edge currents}

\begin{figure}
\scalebox{0.45}[0.45]{\includegraphics[50,427][550,715]{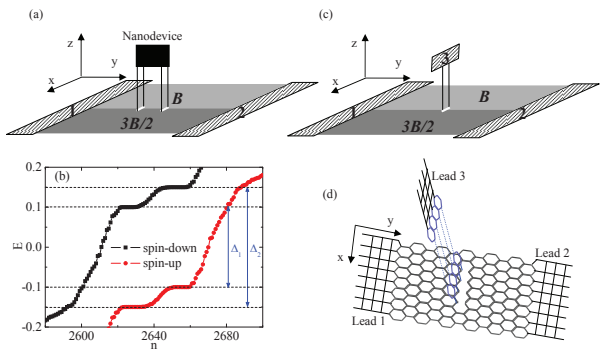}} \caption{\label{fig5}(Color online) (a) Schematic diagram of our proposed
experimental setup for application in spintronics device. (b) The corresponding eigenvalues calculated on the sample in (a) with the magnetic
flux per hexagon $\phi=2\pi/80$ and $1.5\times2\pi/80$ ($B\sim1000$ T). The sample size is $40\times120$ consisting of 40 armchair chains with
120 sites in each chain, while the narrow graphene strip size is $4\times40$. The Zeeman energy is taken as $h_Z=0.1$ and $h_Z=0.15$ for the two
regions respectively. (c) The corresponding three-terminal device with the sample size as (b). (d) Geometry of the graphene sample connected
with three leads for calculating transmission coefficients. To reduce the scattering between the sample and the leads, there exist a transition
region with length $1.14$ nm, where the magnetic field is made to be linearly decreased to zero when approaching leads.}
\end{figure}

So far, we have shown explicitly how to generate and manipulate
the FSP edge currents, but another associated and interesting
question which naturally arises is still unsolved, which is how to
export the FSP current to serve as a current source for a spin
related device. This is not a trivial question, because to export
the current, one has to destroy the interface in some extent to
introduce another lead, but the FSP current is composed of edge
states, which will disappear when the interface is destroyed or
removed away. For this purpose, we design an experimental setup
shown schematically In Fig.~\ref{fig5}(a), which is expected to
export FSP current to the possible spin related device connected
with the armchair graphene sample by a narrow graphene strip. Here
the armchair graphene ribbon is adopted, because for the zigzag
graphene ribbon, the existence of the flat edge bands will cause
ferromagnetic instability and lead to local ferromagnetic
polarizations at sample edges,\cite{Fujita1996} giving rise to
spin-flip processes.

To testify this idea, we first calculate the eigenvalues of this
particular system for an open boundary condition. We find that LLs
are still present even in the condition that the interface is
partially destroyed by the introduced lead. However, states for
both spin species exist between the LLs
$\Delta_1/2<|\mu|<\Delta_2/2$ [Fig.~\ref{fig5}(b)], leaving the
question whether spin-up electrons in this regime contribute to
the current flowing through the device still unaddressed.
Therefore, in order to confirm that the current flowing through
the device is FSP, we refer to the three-terminal setup as
illustrated in Fig.~\ref{fig5}(c), where the rectangular graphene
sample is connected with three ideal semi-infinite leads. The
detail of these connections between graphene sample and the leads
is shown in Fig.~\ref{fig5}(d). In this situation, four bonds
across the interface are broken, so that lead 3 made of the narrow
graphene strip can be connected with the sample. This strip is
then upright so as to avoid the influence of magnetic field.
According to the Fisher-Lee relation, the electron transmission
coefficient from lead $q$ to lead $p$ for multi-terminal
Landauer-B\"{u}ttiker formula is given by\cite{Hankiewicz2004,Sheng2005}
\begin{equation}
T_{pq}^{\sigma}=\texttt{Tr}[\Gamma_p^\sigma G^R\Gamma_q^\sigma G^A], \label{eq:two}
\end{equation}
The scattering rate matrix $\Gamma_p^\sigma$ due to the coupling
to lead $p$ has a form
\begin{equation}
\Gamma_p^\sigma(i,j)=i[\Sigma_{p,\sigma}^R(i,j)-\Sigma_{p,\sigma}^A(i,j)],
\label{eq:three}
\end{equation}
where $\Sigma_{p,\sigma}^R=[\Sigma_{p,\sigma}^A]^\dag$ is the
self-energy due to lead p and the elements of the matrix
$\Sigma_{p,\sigma}^R$ are given by
\begin{eqnarray}
\Sigma_{p,\sigma}^R(i,j)=-\sum_m \chi_m(i)\chi_m(j)e^{ik_ma},
\label{eq:four}
\end{eqnarray}
with the function $\chi_m(i)=\sqrt{\frac{2}{N+1}}\sin\frac{m\pi
i}{N+1}$ describing the transverse profile of mode $m$ in lead $p$
and $k_m$ the wave vector determined by the equation $\mu=-2\cos
k_m-2\cos\frac{m\pi}{N+1}$ . Here $N$ is the number of sites
connected with the leads and $m=1,2,\cdots,N$. The retarded
Green's function of the graphene sample $G^R$ is written as
\begin{equation}
G^R=[EI-H_0-\Sigma^R]^{-1}, ~~~~~ \Sigma^R=\sum_{p,\sigma}\Sigma_{p,\sigma}^R
\label{eq:five}
\end{equation}
and $G^A=[G^R]^\dag$. Here, to make the calculation simple, the
three leads are assumed to be described by a nearest-neighbor
tight-binding model on a square lattice with the same hopping
integral as graphene.

\begin{figure}
\scalebox{0.7}[0.7]{\includegraphics[107,305][414,731]{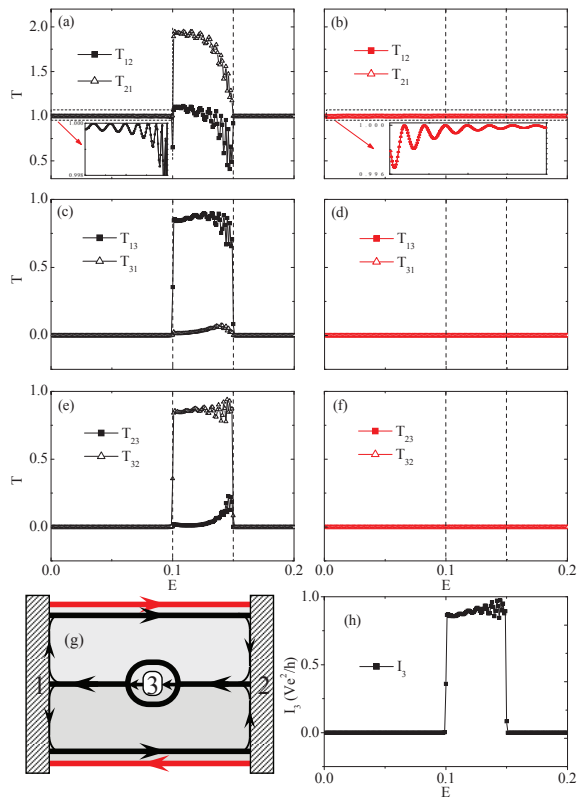}}
\caption{\label{fig6}(Color online) (a)-(f)The transmission
coefficients as functions of the Fermi energy. The coefficients in
(a), (c) and (e) are for spin-down electrons, while the others for
spin-up electrons. The dash lines at $E=0.1$ and $0.15$ indicate
the positions of two spin gaps. (g) Schematic diagram of electron
current flow when $\Delta_1/2<\mu<\Delta_2/2$ (i.e.,
$0.1<\mu<0.15$). (h) The current following out of terminal 3,
$I_3$, when terminal voltages are chosen as $V_1=V_2=V$ and
$V_3=0$. The inset in (a) is an enlarged view of
$T_{12}^\downarrow$ and $T_{21}^\downarrow$ in the gap
$0<\mu<\Delta_1/2$, and the inset in (b) is an enlarged view of
$T_{12}^\uparrow$ and $T_{21}^\uparrow$ in the energy range
$0<E<0.2$.}
\end{figure}

The transmission coefficients $T_{pq}^\sigma$ are plotted in
Fig.~\ref{fig5}. For spin-up electrons, $T_{12}^\uparrow$ and
$T_{21}^\uparrow$ show quasi-periodic oscillation and the maxima
of them are the quantized value 1.  This oscillating behavior is
similar to that found in graphene-based ferromagnetic double
junctions,\cite{Bai2010}and can be attributed to electrons'
multiple reflections between terminal 1 and 2. As the energy
increases gradually from 0 to 0.2, both $T_{12}^\uparrow$ and
$T_{21}^\uparrow$ are approaching to 1, suggesting one left-moving
spin-up edge state on one sample boundary and one right-moving one
on the other, sharing the similar picture with Fig.~\ref{fig2}(a).
However, most importantly, we find that within the error range,
$T_{13}^\uparrow=T_{31}^\uparrow=0$ and
$T_{23}^\uparrow=T_{32}^\uparrow=0$ in the whole energy range
$0<E<0.2$ we are interested in. This is corresponding to the
absence of spin-up edge states at the interface.

For spin-down electrons, when $0<\mu<\Delta_1/2$ and
$\mu>\Delta_2/2$, $T_{12}^\downarrow\approx T_{21}^\downarrow$,
the situation is similar to that for spin-up electrons. However,
when $\Delta_1/2<\mu<\Delta_2/2$, $T_{12}^\downarrow$ and
$T_{21}^\downarrow$ show strongly fluctuations, and deviate from
the quantized value 2 [Fig.~\ref{fig6}(a)], which should be taken
in the ideal case shown in Fig.~\ref{fig2}(a). This may be
ascribed to the bonds breaking at the interface of terminal 3, and
lattice mismatching at the interfaces of terminal 1and 2 which
results in electrons scattering from edges states to the nearby
zeroth LL [see Fig.~\ref{fig5}(b)]. In the same range, the fact
that $T_{13}^\downarrow>T_{31}^\downarrow>0$ and
$T_{32}^\downarrow>T_{23}^\downarrow>0$ indicates that both the
left- and right-moving currents exist, but the net current is left
moving. The finiteness of $T_{23}^\downarrow$ also implies that a
small fraction of spin-down electrons incoming from terminal 3
have to be scattered to sample boundaries via bulk states, since
there is no current channel at the interface from terminal 3 to 2.
Similarly, the finiteness of $T_{31}^\downarrow>0$ means that the
right-moving spin-down electrons propagating along sample
boundaries have some probability to be scattered to terminal 3 via
bulk states. These processes are schematically shown in
Fig.~\ref{fig6}(g). We remark that although these scattering
processes will cause resistance, spin is conserved since there is
no spin-flip processes considered here. This leads to the
conclusion that the electron current flowing out of terminal 3 has
spin-down polarization, and so is FSP.

In actual application, one can apply the electrical voltage to the three terminals as $V_1=V_2=V$ and $V_3=0$, then the FSP (spin-down)
electrons will be transported from lead 3 to leads 1 and 2. In this situation, the FSP current flowing out of terminal 3,
$I_3=(T_{31}^\downarrow+T_{32}^\downarrow)Ve^2/h$ is estimated as $0.85Ve^2/h\sim0.95Ve^2/h$ [see Fig.~\ref{fig6}(h)]. Indeed, the width of the strip is
$d=0.87$~nm and the magnetic length is about $l_B=0.8$~nm. The fact of $d\sim l_B$ guarantees that there are sufficient electrons through the
strip. Therefore, the graphene sample with the designed setup under a NMF can be viewed as the FSP current source for a spin related nanodevice.

\section{summary}
In summary, we have numerically investigated the FSP edge currents
at the interface in graphene under a nonuniform magnetic field.
The manipulation of the FSP currents is realized by a gate
voltage, and its potential application in spintronic devices is
also discussed. An experimental setup is designed and is expected
to export the FSP current, which is confirmed numerically by a
three-terminal transmission calculation.

\begin{acknowledgments}
The author J. An thanks X. G. Wan, H. H. Wen, J. X. Li, Q. H. Wang
for useful discussions. This work was supported by NSFC Projects
No. 10504009, and No. 10874073, and 973 Projects No. 2006CB921802,
No. 2006CB601002, and No. 2009CB929504.
\end{acknowledgments}


\end{document}